\documentclass[10pt,preprintnumbers,aps,amssymb,nofootinbib,amsmath,prl,twocolumn]
{revtex4}
\usepackage{epsfig,epsf}
\usepackage{graphicx}
\usepackage{verbatim}
\usepackage{epstopdf}
\usepackage{bm} 
\newcommand{\beq}{\begin{equation}}
\newcommand{\beql}[1]{\begin{equation}\label{#1}}
\newcommand{\eeq}{\end{equation}}
\newcommand{\bea}{\begin{eqnarray}}
\newcommand{\eea}{\end{eqnarray}}
\def\eq#1{{(\ref{#1})}}
\def\fig#1{{Fig.~\ref{#1}}}

\newcounter{topiccounter}
\def\b#1{\mathbf{#1}}

\newcommand{\im}{\mathrm{Im}\,}

\begin{document}

\preprint{RBRC-813}

\title{Geometry of attosecond laser pulses and photon-photon scattering at high energies }

\author{Kirill Tuchin$\,^{a,b}$\\}

\affiliation{
$^a\,$Department of Physics and Astronomy, Iowa State University, Ames, IA 50011\\
$^b\,$RIKEN BNL Research Center, Upton, NY 11973-5000\\}

\date{\today}


\begin{abstract}
We derive the total cross section for scattering of a photon on an ultra-short  laser pulse at high energies. We take into account all multi-photon interactions.  
We argue that  the nonlinear effects due to these interactions become important at very high intensities of the laser pulse. We demonstrate however, that these intensities are significantly lower than the Schwinger critical value. 
\end{abstract}

\maketitle

Geometry of a laser pulse $\ell$ plays a prominent role in its interactions with elementary particles \cite{Baier:1989ar,Tuchin:2009sg}.  In this letter we consider interactions  of photons with ultra-short laser pulses at high energies. Laser pulse is a coherent state of  large number of photons $N$. If  width $2d$ and length $L$ of the laser pulse can be neglected, then the total $\gamma\ell$ cross section reads \cite{Lipatov:1969sk,Cheng:1970ef}
\beql{if}
\sigma_\mathrm{tot}^{\gamma\ell}(s)= \frac{\alpha^4 N}{ m^2}\frac{1}{36\pi}[175\zeta(3)-38]\,.
\eeq
This is just $N$ times the total  $\gamma\gamma$ cross section, which at high energies receives the leading contribution from the process $\gamma\gamma\to e^-e^+e^-e^+$ mediated by the Weisz\"acker-Williams photon  exchanged in the $t$-channel between the two $e^+e^-$ electric dipoles. At high intensities $I$ of the laser pulse, i.e.\ at large $N$, close to the Schwinger critical value,  the non-perturbative effects as well as non-linear perturbative ones are expected to produce large corrections to Eq.~\eq{if}. In this letter we demonstrate that it is in principle possible to prepare a laser pulse that would interact non-linerly with energetic photons at intensities smaller than the critical. 

We begin with an observation that  the laser pulse width $2d$ is actually  
 large compared to the effective radius of the electro-magnetic interactions. Indeed, the maximal impact parameter between the two $e^-e^+$ pairs is $b'_\mathrm{max}= \frac{\sqrt{s}}{4m^2}$, which is the consequence of existence  of the minimal longitudinal momentum transfer \cite{Berestetsky:1982aq} and the fact that the $t$-channel photon is almost real $l^2\approx -\b l^2$ (we use bold face for transverse -- w.r.t.\ the collision axes -- two-vectors). Here $s$ and $t$ are the usual Mandelstam variables.  For realistic values of $s$ and $d$ it holds that $b'_\mathrm{max}\ll d$. For example, consider collision of $\omega=100$~GeV photon with $\Omega=100$~eV laser pulse. The high energy approximation holds since  $\sqrt{s}=6.3\gg 1$~MeV. The pulse width can be estimated as $d\sim \lambda=2\pi/\Omega=1.3$~nm; hence   $b'_\mathrm{max}/d=6\cdot 10^{-3}\ll 1$. We will often refer to this example.

To take the finite width of the laser pulse into account it is convenient to Fourier transform the amplitude of  $\gamma\ell$  scattering into the transverse configuration space. Then, the wave-function  $\tilde\Phi(\b r)$ of the energetic photon, describing splitting of a photon into $e^-e^+$ dipole, factorizes from the  imaginary part of elastic scattering amplitude $i\Gamma^{d\ell}$, where $d$ stands for the electric dipole  $e^-e^+$ produced by the energetic photon.
Let $\b b'$ be the impact parameter between the dipole $d$ and a photon $\gamma_\ell$ in the pulse; $\b B$ -- between the dipole $d$  and the pulse symmetry axes and $\b b$ -- between the photon $\gamma_\ell$  and the symmetry axes; clearly, $\b b'=\b B-\b b$.  The total cross section reads \cite{Donnachie:1999kp,Bondarenko:2003yma}
\beql{gmp-l}
\sigma^{\gamma \ell}_\mathrm{tot}(s)=\frac{1}{2!}\int d^2B\int \frac{d^2r}{2\pi} \,\tilde\Phi(\b r)\,
 2\,\langle\im [i\Gamma^{d\ell}( \b r,\b B, s)]\rangle\,,
\eeq
where \cite{Nikolaev:1990ja}
\beql{phi-z}
\tilde \Phi(\b r)= \frac{\alpha m^2}{\pi}\bigg( \frac{2}{3}K_1^2(m r)+K_0^2(m r)\bigg)\,.
\eeq

The longitudinal extent $L$ of the laser pulse is also an important parameter. It is determined by the pulse duration $\tau$. For an attosecond pulse $\tau\sim   10^{-18}$~sec which translates into $L\sim \tau c= 0.3$~nm. For highly energetic photons $L$ may turn out to be shorter than the coherence length $l_c$ of the energetic photon\footnote{This is not to be confused with the laser coherence length.}, which is given by the inverse of the minimal longitudinal momentum transfer. In the Lab frame \cite{Berestetsky:1982aq} 
\beql{coh.len}
l_c= (\hbar\omega/2mc^2)\lambdabar\,,
\eeq
where $\lambdabar=3.8\cdot 10^{-4}$~nm is the Compton wavelength of electron. 
Coherence effects due to multiple scattering of an energetic photon on photons  of the laser pulse are important when $l_c\gg L$ which implies a condition 
\beql{maxL2}
\frac{\hbar\omega}{2mc^2}\gg \frac{L}{\lambdabar}\,.
\eeq
In the following we assume that \eq{coh.len} is satisfied. In particular, for a sample parameter set chosen above $l_c/L\simeq 200$.  

\begin{figure}[ht]
      \includegraphics[width=7cm]{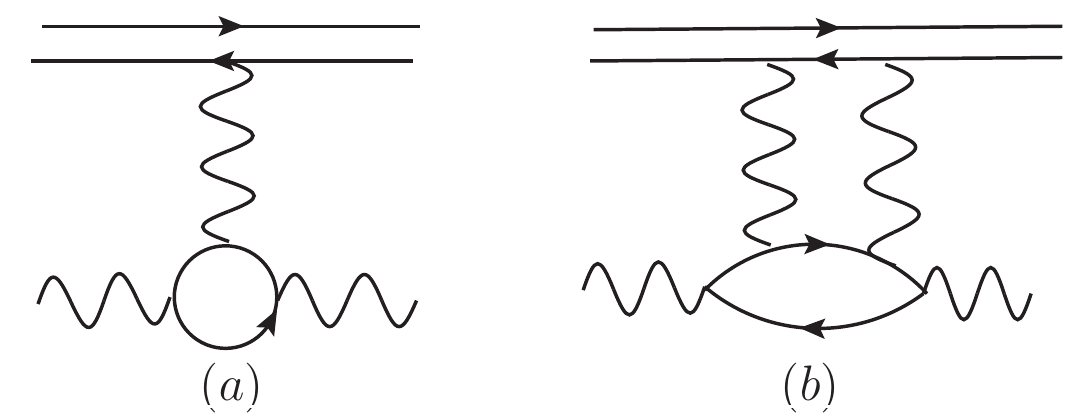} 
  \caption{Leading order contributions to the dipole-photon elastic cross section. The upper dipole is associated with the energetic photon. The horizontal wavy line is  a photon in the laser pulse.}
\label{fig:gamma}
\end{figure}

Since all $N$ photons in the pulse are in a coherent state we can write using the Glauber approach \cite{Glauber:1987bb}
\beql{indep}
\langle\im [i\Gamma^{d\ell}( \b r,\b B, s)]\rangle= \im (1-e^{-N\langle i\Gamma^{d\gamma_\ell}( \b r,\b B, s)\rangle})\,,
\eeq
where $i\Gamma^{d\gamma_\ell}$ is elastic scattering amplitude of the dipole $d$ on a photon $ \gamma_\ell$.
Averaging in \eq{gmp-l} and \eq{indep} is performed over the entire volume of the pulse. 
At high energies, the interaction is approximately instantaneous (since the mediating photons are almost real) implying that the dipoles in the pulse do not recoil. One then describes the number distribution of the dipoles in the pulse by a function $\rho(\b b, z)$, where $z$ is the longitudinal position of a dipole. It is normalized such that  
\beq\label{dens.norm-l}
\int d^2b\, dz\,\rho(\b b,z)= N\,.
\eeq
The average of the amplitude over the dipole position in the pulse reads
\beql{a1-l}
\langle i \Gamma^{d \gamma_\ell}(\b r, \b B, s)\rangle= \frac{1}{N}\int d^2b\int_{-L/2}^{L/2} dz\, \rho(b)\,i \Gamma^{d \gamma_\ell}(\b r, \b B-\b b, s).
\eeq
We estimated above that $b_\mathrm{max}'\ll d$. Since $b \sim d$ it implies that $b'\ll B\approx b$. Neglecting for notational simplicity $z$-dependence of $\rho$ we write
\beql{a3}
\langle i \Gamma^{d \gamma_\ell}(\b r, \b B, s)\rangle\approx \frac{L}{N}\,\rho(B) \int d^2b' i \Gamma^{d \gamma_\ell}(\b r,\b b', s)\,.
\eeq
To the leading order in $\alpha$ the dipole-photon elastic scattering amplitude is depicted in \fig{fig:gamma}. We can write analogously to   \eq{gmp-l} 
\beql{tt}
\im  [i\Gamma^{d \gamma_\ell}(\b r,\b b', s)]= 
\int \frac{d^2r'}{2\pi}\tilde \Phi(r')\,\im[i \Gamma^{dd}(\b r,\b r',\b b',s)]\,,
\eeq
where $i \Gamma^{dd}$ is the elastic dipole--dipole scattering amplitude.

Diagram \fig{fig:gamma}(a) represents the leading contribution to the real part of the dipole-photon scattering amplitude. However, it vanishes due to the $C$-invariance of QED (Furry's theorem). Higher order contributions to the real part of the dipole--photon amplitude are represented by diagrams with odd number of exchanged photons. They vanish for the same reason. Therefore, the  dipole-photon amplitude is purely imaginary at high energies. The leading diagram is shown in \fig{fig:gamma}(b). The corresponding dipole--dipole amplitude is 
\beql{sc.ampl-l2}
i\Gamma^{dd}= 2i \alpha^2\ln^2\frac{|\b b'+\frac{1}{2}\b r+\frac{1}{2}\b r'||\b b'-\frac{1}{2}\b r-\frac{1}{2}\b r'|}{|\b b'+\frac{1}{2}\b r-\frac{1}{2}\b r'||\b b'-\frac{1}{2}\b r+\frac{1}{2}\b r' |} \,.
\eeq
Using \eq{gmp-l},\eq{indep},\eq{a3},\eq{tt},\eq{sc.ampl-l2} we derive for the total pair-production cross section
\beql{sig}
\sigma_\mathrm{tot}^{\gamma\ell}= \frac{\alpha}{\pi}\int d^2B\int_0^\infty
du  u\, \tilde\Phi(u)\big\{ 1-  e^{-\kappa(u,B)}\big\}  \,,
\eeq
where 
\beql{kk}
\kappa = \frac{8\alpha^3\rho(B)\, L}{m^2}u^4
\int_0^1 d\xi \log\frac{e}{\xi}\big[ \xi^3 \tilde \Phi(u\xi)+\xi^{-3}\tilde \Phi(u\xi^{-1})
\big]
\eeq
with $\xi=r'/r$ and $u=m r$.

When $\kappa=N\langle i\Gamma^{d\gamma_\ell}\rangle\ll 1$ the exponent in \eq{indep} and \eq{sig} can be expanded which corresponds to the linear two-photon exchange regime. Integration over dipole sizes and impact parameters then yields \eq{if}. In the opposite limit $\kappa\gg 1$ the nonlinear effects are strong leading to saturation of the cross section at its geometric limit $\sigma_\mathrm{tot}^{\gamma\ell}\sim \alpha\, d^2\ln \kappa$. This regime is characterized by weak logarithmic dependence of the cross section on intensity of the laser pulse. Given the laser pulse of wavelength $\lambda$ and pulse duration $\tau$,
transition from the linear to the saturation regime can be characterized by intensity $I_\mathrm{mp}$ for which $\kappa\simeq 1$. We can estimate 
\beql{kest}
\kappa\simeq \frac{8\alpha^3\tau\lambda\lambdabar^2 I}{\pi\hbar c}\,.
\eeq
Therefore
\beql{Int-sat}
I_\mathrm{mp}=\frac{\pi \hbar c}{8\alpha^3\tau \lambda \lambdabar^2}\,.
\eeq

Let us compare $I_\mathrm{mp}$ with the critical value $I_c$ at which the Schwinger mechanism of pair production \cite{Sauter:1931zz,Heisenberg:1935qt,Schwinger:1951nm} becomes important. $I_c=cE^2_c$ where the critical value of electric field is $E_c\sim m^2c^3/(e\hbar)$. Therefore,
\beql{comp}
\frac{I_\mathrm{mp}}{I_c}\sim \frac{1}{\alpha^2} \frac{\lambdabar^2}{L\lambda}\ll 1
\eeq
for realistic values of $L$ and $\lambda$. In fact, for the sample set of parameters that we are using  $I_\mathrm{mp}=7\cdot 10^{-5}\, I_c$.

In conclusion, we argued that  the geometry of attosecond laser pulses is such that at presently accessible high energies
\texttt{(i)}  the  width of the pulse $d$ is much larger than the effective radius  electro-magnetic interactions, \texttt{(ii)} length of the pulse $L$  is shorter than the coherence length. For this case we derived the total cross section \eq{sig} that includes all multi-photon interactions between the energetic photon and the laser pulse. These multi-photon interactions are important at laser intensities $I_\mathrm{mp}$ much smaller than the critical value at which the perturbative vacuum of QED becomes unstable. This result may be useful for investigation of  nonlinear effects in electrodynamics (see e.g.\ \cite{Marklund:2006my}).

\bigskip
\acknowledgments
This work  was supported in part by the U.S. Department of Energy under Grant No.\ DE-FG02-87ER40371. I 
thank RIKEN, BNL, and the U.S. Department of Energy (Contract No.\ DE-AC02-98CH10886) for providing facilities essential
for the completion of this work.


\end{document}